\documentclass[aps,prd,preprint,tightenlines,floatfix,showpacs,
groupedaddress,superscriptaddress]{revtex4}

\usepackage{graphicx}
\usepackage{subfigure}

\begin{document}
\preprint{UFIFT-HEP-07-17}
\preprint{BI-TP 2007/34}

\vspace{1cm}

\date{December 20, 2007}

\title{A further look at particle annihilation in dark matter caustics}

\author{Aravind Natarajan}
\email{anatarajan@physik.uni-bielefeld.de}
\affiliation{Institute for Fundamental Theory, Department of Physics,
   University of Florida, Gainesville, FL  32611-8440, U.S.A.}
 \affiliation{Fakult\"{a}t f\"{u}r Physik, Universit\"{a}t Bielefeld, Universit\"{a}tsstra$\beta$e 25, Bielefeld D-33615, Germany}
 
\author{Pierre Sikivie}
\email{sikivie@phys.ufl.edu}
\affiliation{Institute for Fundamental Theory, Department of Physics,
   University of Florida, Gainesville, FL  32611-8440, U.S.A.}

\begin{abstract} 

Dark matter caustics are small scale, high density structures believed to exist in galaxies like ours. If the dark matter consists of Weakly Interacting Massive Particles, these caustics may be detected by means of the gamma rays produced by dark matter particle annihilation. We discuss particle annihilation in outer and inner caustics and provide sky maps of the expected gamma ray distribution.

\vspace{0.34cm}
\noindent
\end{abstract}
\maketitle

\section{Introduction \label{section1}}

A consequence of the presence of cold dark matter in galactic 
halos is the formation of caustics. Caustics are regions of 
infinite density in the limit where the dark matter particles 
have zero velocity dispersion. The leading dark matter candidates, 
namely axions and Weakly Interacting Massive Particles(WIMPs), have 
very small primordial velocity dispersions ($\sim 0.03$ cm/s for a 
100 GeV WIMP, $\sim 10^{-6}$ cm/s for a 10 $\mu$eV axion \cite{ps_crs})  
and may be expected to form caustics with large density enhancements. 
If the dark matter is made up of WIMPs, the particles can annihilate
producing gamma rays. It may be possible to detect this gamma ray 
flux by means of future experiments such as GLAST \cite{glast}. 
Caustics may possibly be revealed by this technique.

Caustics form because cold dark matter particles occupy a thin, three 
dimensional hypersurface in phase space \cite{ps_crs,caustic_papers}. 
There are two kinds of caustics in galactic halos: outer and inner
\cite{ps_crs}.  The outer caustics are topological spheres surrounding
galaxies. The inner caustics have a more complicated geometry which
depends on the details of the dark matter angular momentum distribution
\cite{ns2}. When the angular momentum distribution is dominated by a 
net rotational term, the inner caustics have the appearance of rings
\cite{ps_crs,ns2}. Observational evidence for caustic rings was found 
by analyzing the rotation curves of external galaxies \cite{ks} and 
the rotation curve of the Milky Way \cite{iras}. A triangular feature 
seen in the Infrared Astronomical Satellite (IRAS) map of the Milky Way 
is also suggestive of the presence of a nearby inner caustic ring
\cite{iras}. Recently \cite{os}, possible evidence was found for a 
caustic ring in the galaxy cluster Cl0024+1654.  In this article, we
discuss dark matter annihilation in caustic spheres and caustic rings 
and present simulated gamma ray sky maps.

The existence of dark matter caustics is sometimes disputed on the
basis of large scale cosmological simulations. However, current
simulations use particles with mass of order $\sim 10^6 M_\odot$. 
Such massive particles have spurious collisions which destroy fine
details in the dark matter phase space hypersurface. Thus the absence 
of caustics in most large scale simulations is probably due to the finite
resolution of these simulations and the large particle masses involved.
Note that discrete flows and caustics are observed in $N$-body 
simulations when special care is taken.  The simulations of Stiff 
and Widrow \cite{Sti}, which increase the resolution in the relevant
regions of phase space, do show discrete flows.  More recently, 
Vogelsberger et al. \cite{Voge} employed a new technique to study 
small scale structures such as caustics.  Further progress in 
numerical techniques may enable the study of caustic properties 
in galactic halos using $N$-body simulations.

In cold dark matter cosmology, galactic halos form through 
a process of ``hierarchical clustering".  Small halos form
first.  Larger halos, such as that of the Milky Way, are the 
outcome of past mergers, the accretion of smaller halos of 
various sizes, and the accretion of unclustered dark matter.  
This general scenario must surely be correct without, however, 
telling us anything precise.  The critical issue for the existence 
of a caustic is the velocity dispersion of the flow that would 
form it.  To be specific, consider the flow of dark matter 
that the Milky Way halo is accreting {\it today}.  If the
dark matter in this flow has not clustered on any small 
scales yet, its velocity dispersion has the primordial value 
mentioned earlier, of order 0.03 cm/s for a 100 GeV WIMP.  If 
it has clustered already, its effective velocity dispersion is 
the velocity dispersion of the structure it has become part of.  
The largest structures for which we have evidence of recent 
accretion are small sattelites, such as the Sagittarius dwarf, 
with velocity dispersion of order 10 km/s.  So one may conservatively 
assume that the flow of WIMP dark matter accreting onto the Galaxy 
today has variable velocity dispersion between 0.03 cm/s and 10 km/s.  

Dark matter caustics are generically surfaces on one side of which 
the density diverges, in the limit of zero velocity dispersion, as  
\begin{equation}
d(x) = {A \over \sqrt{x}}
\label{foldden}
\end{equation}
where $x$ is the distance to the surface and $A$ is a constant
called the ``fold coefficient".  For finite velocity dispersion 
$\delta v$, the dark matter caustics in a galactic halo are 
smoothed over a distance
\begin{equation}
\delta x \sim {\delta v \over v} R
\label{cutoff}
\end{equation}
where $v$ is the flow velocity (of order 300 km/s) and $R$ is 
the overall size of the halo (of order 300 kpc).  The velocity 
dispersion cuts off the divergence of the density at the caustic, 
i.e. the ${1 \over \sqrt{x}}$ behaviour stops for $x < \delta x$.
For the above range of velocity dispersion, $\delta x$ ranges from
$10^{15}$ cm to 10 kpc.  Since outer caustics have size of order 
hundreds of kpc, they are not erased by the expected velocity 
dispersion.  Inner caustics which have size of order 10 kpc
are smoothed out only when the flow is entirely composed of 
sattelites with velocity dispersion of order 10 km/s.

As was mentioned, evidence  for caustic rings of dark matter 
in the Milky Way was found in the existence and distribution 
of sharp rises in the Galactic rotation curve and the presence 
of a triangular feature in the IRAS map of the Galactic plane 
\cite{iras}.  This evidence implies an upper limit on the 
velocity dispersion of order 50 m/s, or an upper limit on the 
smoothing scale $\delta x$ of order $10^{-4}~R \sim 30$ pc.

The WIMP annihilation rate per unit surface of caustic is 
proportional to the integral 
\begin{equation}
\int dx~d(x)^2 \simeq \int_{\delta x}^R {A^2 \over x}
= A^2 \ln\left({R\over \delta x}\right)~~~\ .
\label{logenh}
\end{equation}
This equation shows that caustics enhance the annihilation 
signal only by a logarithmic factor \cite{detecting} which 
varies from $\sim 3$ to 20, depending on the velocity dispersion.

In this article, we limit ourselves to the simple cases of a 
spherically symmetric outer caustic and an axially symmetric
inner caustic ring.  Our symmetry assumptions are made solely 
for the purpose of computational ease.  In previous work \cite{ns2},
we verified that caustics are stable under perturbations, in 
particular perturbations that break any assumed symmetry.  
Moreover, the symmetry assumptions are consistent with the 
observational evidence mentioned earlier. 

Previous work on the possibility of detecting dark matter caustics by
indirect means include \cite{detecting,pieri,an,mss}. Caustics and cold
flows are also relevant to WIMP and axion direct detection experiments
\cite{direct}.

The number of photons received by a detector per unit area, per unit time,
per unit solid angle and per unit energy due to WIMP annihilation is given
by the expression\cite{pieri,gondolo} 
\begin{equation}
\frac{d\Phi}{d\Omega dE_\gamma }(E_\gamma,l,b) =
\frac{dS}{dE_\gamma}(E_\gamma) \times \frac{E(l,b)}{4\pi}.
\label{wimp_formula} 
\end{equation} 
The term $dS/dE_\gamma$ depends only
on the particle physics: \begin{equation} \frac{dS}{dE_\gamma}(E_\gamma) =
\frac{ <\sigma_a v>}{2 m^2_\chi} \sum_i b_i \;
\frac{dN_{\gamma,i}}{dE_\gamma}(E_\gamma).  \end{equation} $<\sigma_a v>$
is the WIMP annihilation cross section times the relative velocity
averaged over the momentum distribution of the WIMPs, $m_\chi$ is the WIMP
mass, $dN_{\gamma,i}/dE_\gamma$ is the number of photons produced per
annihilation per unit energy $E_\gamma$ in the annihilation channel $i$,
and $b_i$ is the branching fraction of channel $i$.

The term $E(l,b)$ is called the emission measure and depends solely on the
dark matter distribution.
 \begin{equation}
 E(l,b) = \int_{los} dx \; d^2(x)
 \end{equation}

where $x(l,b)$ is the distance measured along the line of sight($los$) in
the direction of the galactic co-ordinates $(l,b)$, and $d$ is the dark
matter density. Since all detectors have a finite resolution, the emission
measure is often expressed in terms of $<E>$, which represents an average
over the resolution of the detector. In this article, we limit ourselves
to computing $E(l,b)$.

\begin{figure}[!h]
\scalebox{0.4}{\includegraphics{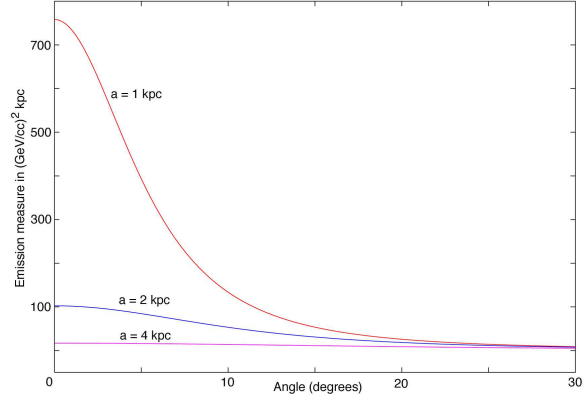}}
\caption{Emission measure due to a smooth halo with the density profile of Eq. \ref{smooth} with $a$ = 1 kpc, 2 kpc and 4 kpc. $r_0 = 8.5$ kpc and $d_0 = 0.3$ GeV/cm$^3$.  \label{figure1}}
\end{figure}

As an example, let us compute the emission measure due to a smooth
isotropic halo with the density profile
\begin{equation}
d(r) = d(r_0) \, \frac{ r_0^2 + a^2 }{ r^2 + a^2 }. \label{smooth}
\end{equation}
Consider an observer at a distance $r_0$ from the center of the halo, and
a line of sight making an angle $\psi$ with the line joining the observer
with the center (we choose $\psi = 0$ to be the direction of the halo
center). $a$ is the core radius and keeps the density finite. We choose $d
= 0.3$ GeV/cm$^3$ when $r = r_0 = 8.5$ kpc.  The emission measure $E$ is
plotted as a function of angle $\psi$ in Fig. \ref{figure1}, for the cases
$a = 1,2$ and 4 kpc.
 
 \section{Outer caustics \label{section2}}
 
Outer caustics are topological spheres. They are fold catastrophes with the universal density profile
\begin{equation}
d(r) = \frac{A}{\sqrt{R-r}} \; \Theta(R-r)
\label{caustic_density}
\end{equation}
where $d(r)$ is the dark matter density, $r$ is the distance from the center of the sphere, measured along the radial direction, $R$ is the radius of the sphere, $A$ is a constant for the particular caustic, and $\Theta$ is the unit step function. For the self-similar infall model \cite{ss1,ss2,ss3} of the Milky Way halo, the radii of the first $n=1,2,3,\cdots$ outer caustics $R_n \simeq \{ 440, 260, 190, 150, 120, \cdots \}$ kpc \cite{ss3} and the corresponding values of  $A_n$ $\simeq \{ 8,9,10,10,11,\cdots \} \times 10^{-4} M_\odot / \textrm{pc}^{5/2}$ \cite{ss3}. We will assume a thickness $\delta R$ for the caustic sphere, which sets a density cutoff equal to $A/\sqrt{\delta R}$.

\subsection{Observer outside the caustic sphere}

\begin{figure}[!h]
\begin{center}
\scalebox{0.4}{\includegraphics{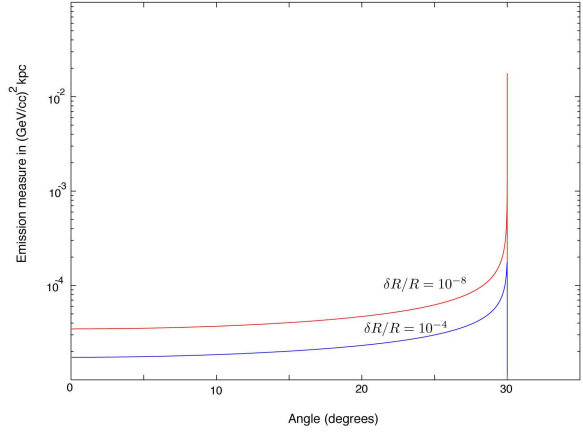}}
\end{center}
\caption{Emission measure from a spherical outer caustic(observer outside the sphere) for $\delta R/R = 10^{-8}$ and $\delta R/R = 10^{-4}$. $r_0/R = 2$.  The constant $A$ was chosen to be $0.97 \times 10^{-3} \frac{ \textrm{GeV} }{ \textrm{cc} } \sqrt{ \textrm{kpc} } = 8.1 \times 10^{-4} \frac{ \textrm{M}_\odot } { \textrm{pc}^{5/2} }$. \label{figure2}}
\end{figure}

\begin{figure}[!h]
\begin{center}
\scalebox{0.8}{\includegraphics{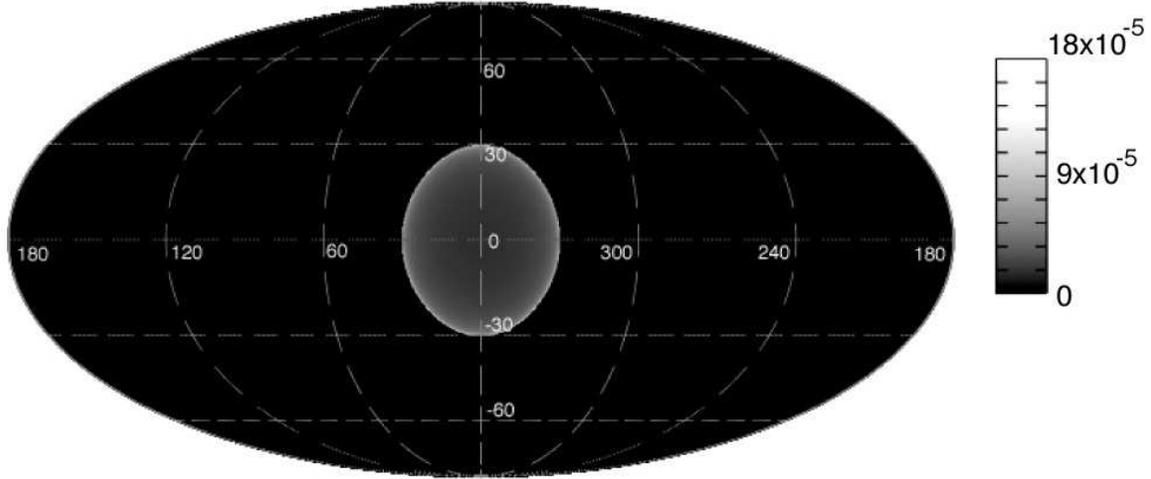}}
\end{center}
\caption{Simulated skymap of the emission measure from a spherical outer caustic, with the observer being located outside the sphere. $r_0/R = 2, \delta R/R = 10^{-4}, A = 0.97 \times 10^{-3} \frac{ \textrm{GeV} }{ \textrm{cc} } \sqrt{ \textrm{kpc} }$. The scale shows $E$ / (GeV/cc)$^2$ kpc. \label{figure3}}
\end{figure}

Consider a spherical outer caustic of radius $R$ and thickness $\delta R$. Let an observer be located at a distance $r_0$ from the center of the sphere $(r_0 > R)$. Consider a line of sight passing through the sphere making an angle $\psi$ with the line joining the observer with the center of the sphere($\psi = 0$ is the direction of the center). The emission measure $E$ is plotted in Fig. \ref{figure2} for $r_0 / R = 2$, for the two cases $\delta R / R = 10^{-8}$ and $\delta R / R = 10^{-4}$.  The maximum value of $E \approx 1.87 A^2 \sqrt{R/\delta R}$ occurs when the line of sight is tangent to the sphere.  Fig. \ref{figure3} shows a simulated sky map. 

\subsection{Observer inside the caustic sphere}

\begin{figure}[!h]
\begin{center}
\scalebox{0.7}{\includegraphics{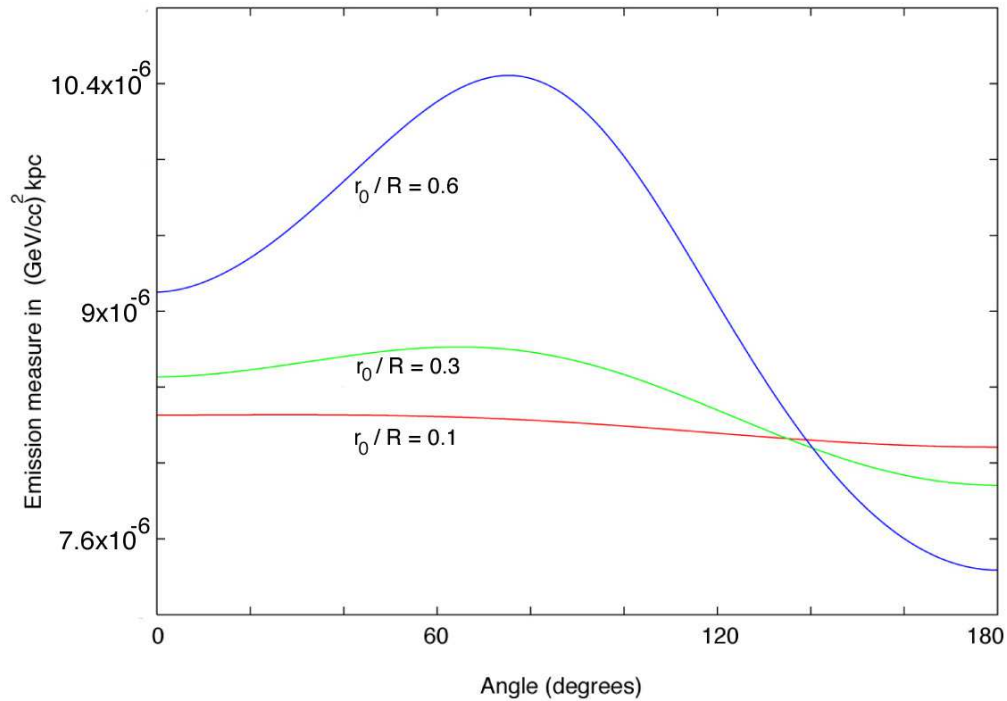}}
\end{center}
\caption{Emission measure from a spherical outer caustic (observer inside the sphere). Shown are three cases: $r_0/R=0.1,0.3,0.6$. The constant $A$ was chosen to be $0.97 \times 10^{-3} \frac{ \textrm{GeV} }{ \textrm{cc} } \sqrt{ \textrm{kpc} }$ and $\delta R/R = 10^{-4}$.  \label{figure4}}
\end{figure}

\begin{figure}[!h]
\begin{center}
\scalebox{0.6}{\includegraphics{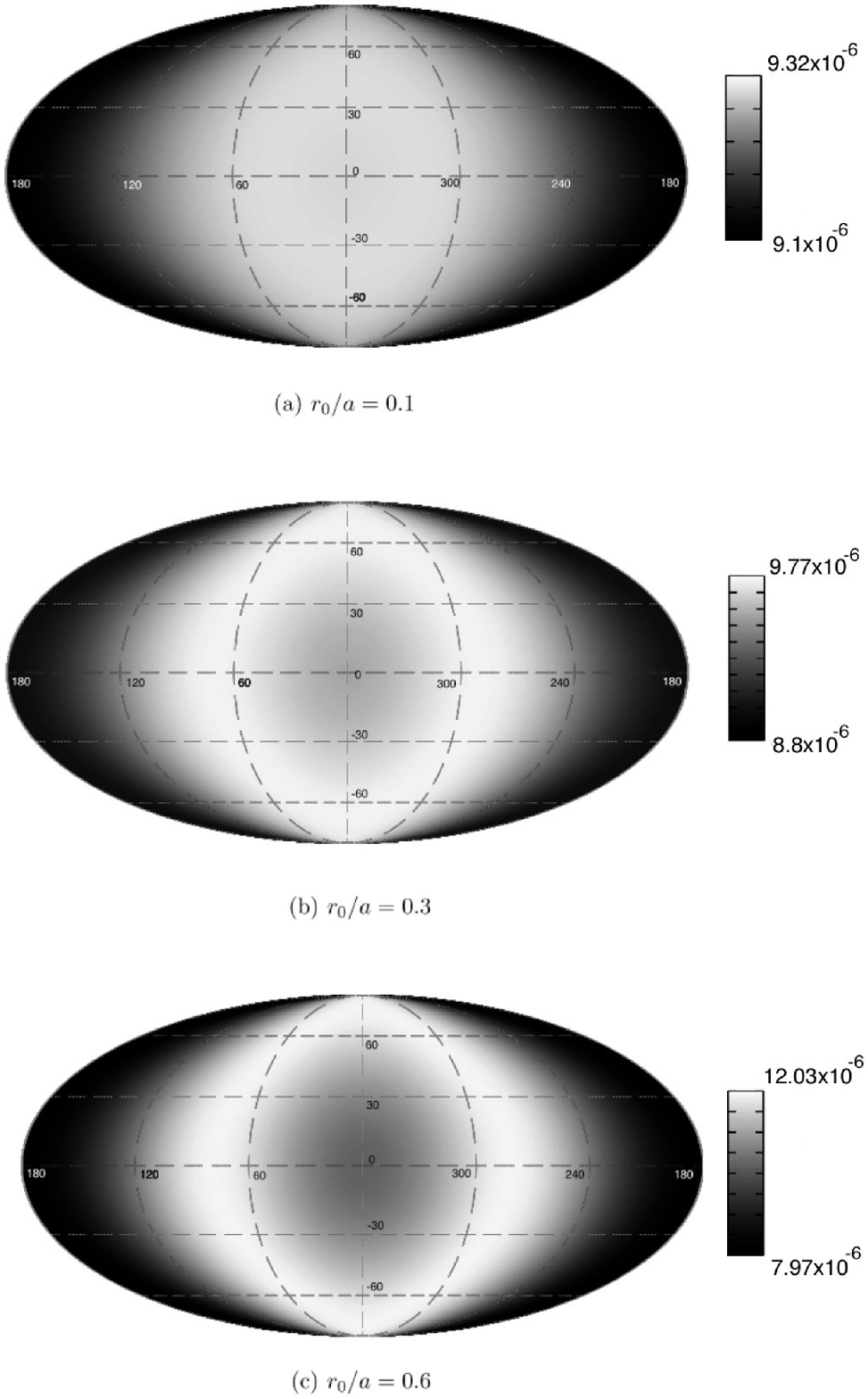}}
\end{center}
\caption{Simulated sky maps of the emission measure due to a spherical outer caustic, with the observer located within the sphere. Shown are the three cases: $(a) r_0/R = 0.1, (b) r_0/R = 0.3$ and $(c) r_0/R = 0.6$. $A$ was chosen to be $0.97 \times 10^{-3} \frac{ \textrm{GeV} }{ \textrm{cc} } \sqrt{ \textrm{kpc} }$. $\delta R/R = 10^{-4}$. The scale shows $E$ / (GeV/cc)$^2$ kpc. \label{figure5}}
\end{figure}

Let us now consider the case where the observer is within the caustic sphere. We once again assume a sphere of radius $R$ and thickness $\delta R$. The observer is located at a distance $r_0$ from the center($r_0 < R$). As before, $\psi$ is the angle made by the line of sight with the line joining the observer with the center ($\psi=0$ gives the direction of the center). Fig. \ref{figure4} shows the emission measure plotted as a function of angle $\psi$ for the three cases  $r_0/R = 0.1,0.3$ and $0.6$. Figs. \ref{figure5} (a),(b),(c) show simulated skymaps.

In both cases, the emission measure for the first few outer caustics of a halo like the Milky Way is quite small compared to that obtained from the smooth halo of Eq. \ref{smooth}, for $\delta R/R > 10^{-10}$. However, $E$ would be larger for caustics of smaller radii and could possibly make a significant contribution to the total flux \cite{mss}.

\section{Inner caustics \label{section3}}

The inner caustics are made up of sections of the higher order catastrophes. As mentioned in the Introduction, the inner caustics resemble rings when the dark matter angular momentum distribution is dominated by a net rotational term. Here, we assume that this is the case. A caustic ring is a closed tube whose cross section has three cusps. Consider a caustic ring of radius $a$, horizontal extent $p$ and vertical extent $q$. See \cite{ps_crs} for details. We choose cylindrical co-ordinates $(z,\rho=\sqrt{x^2+y^2})$. Let us define the two dimensionless variables $X = (\rho-a)/p$ and $Z = z/q$. For simplicity, we assume axial symmetry about the $Z$ axis and reflection symmetry about the $Z=0$ plane. With these assumptions, we can obtain an analytic expression for the density $d$ close to a caustic ring, in the limit $p,q \ll a$. For the special case of $Z=0$, the dark matter density is given by \cite{an}
\begin{equation}
d(X,0)  =  \frac{f v^2_{rot}}{4 \pi G}   \left \{
    \begin{array}{lll}
         \frac{1}{1-X}   &    \mbox{ when $X \le 0$   } \\
         \frac{1}{1-X} \left( 1 + \frac{1}{\sqrt{X}} \right )  &  \mbox{ when $0 \le X \le 1$ } \\
         \frac{1}{X-1}   &    \mbox{ when $X \ge 1$.   } 
     \end{array}
  \right.
  \label{d1}
\end{equation}
$f$ is a constant for the particular caustic ring. For the self-similar infall model of the Milky Way halo \cite{ss2,ss3}, $f_n \approx  \{0.11, 0.046, 0.029, 0.021, 0.017,...\}$ for the first few $n=1,2,3,...$ dark matter flows. For the more general case of non-zero Z, the density $d(Z,X^2)$ takes the form \cite{an}:
\begin{equation}
d(X,Z^2) = \frac{f v^2_{rot}}{4 \pi G}  \;  \sum_i \frac{1}{ 2 \left| 1 - X + 2\,T^2_i - 3 T_i  \right | }
\end{equation}
where $T_i$ are the real roots of the quartic equation
\begin{equation}
T^4 - 2\,T^3 + (1-X)\,T^2 - \frac{27}{64} Z^2 = 0.
\label{T}
\end{equation}
Note that the assumption of reflection symmetry about the $Z=0$ plane implies that the dark matter density is a function of $Z^2$. See Appendix \ref{app} for a derivation of Eqs. \ref{d1} - \ref{T}

\subsection{Observer outside the caustic ring tube}

\begin{figure}[!h]
\begin{center}
\scalebox{1.0}{\includegraphics{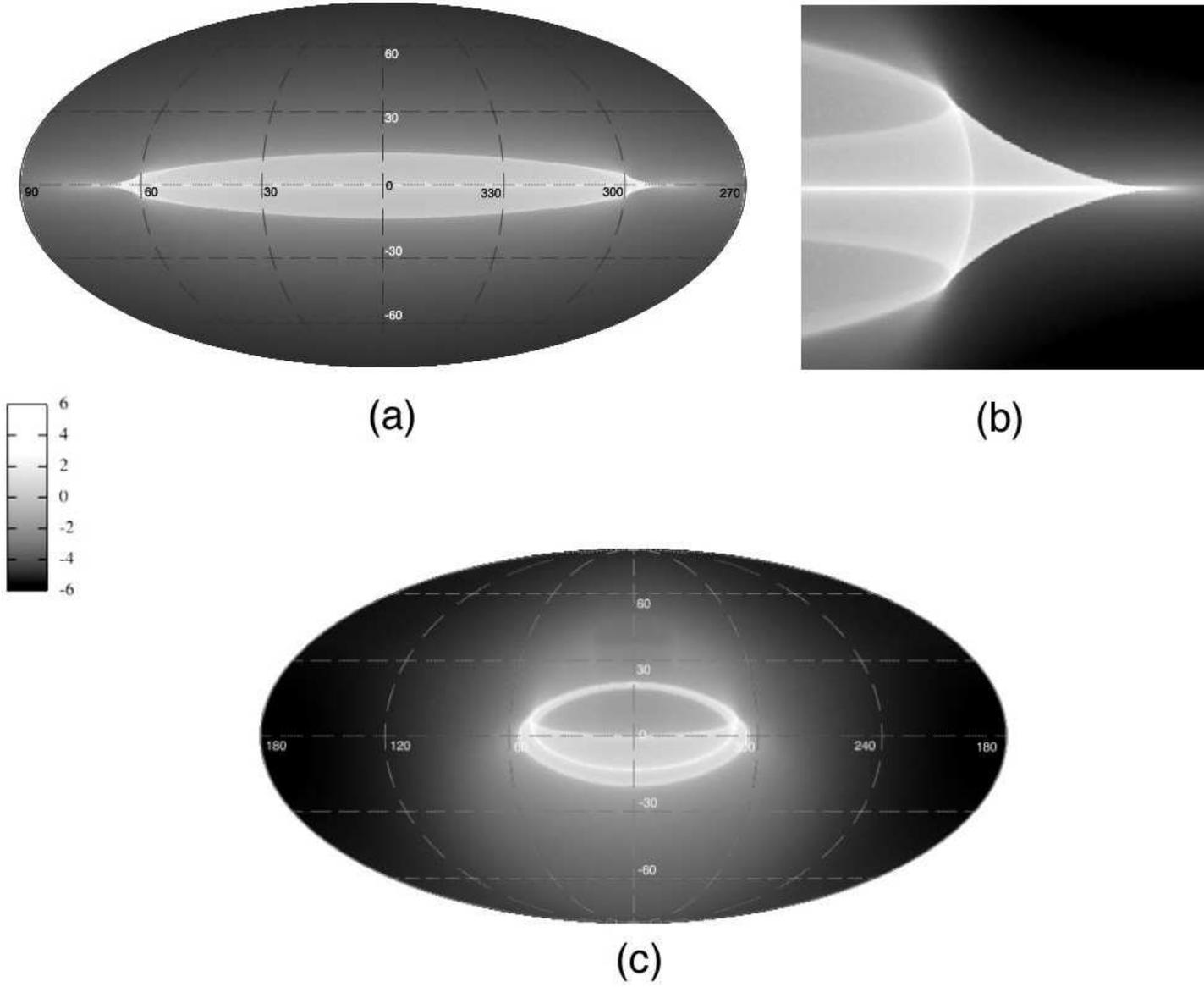}}
\end{center}
\caption{Simulated sky maps of the emission measure from an inner caustic ring, with the observer located outside the ring. $f$ was chosen to be 0.02 and $p/q = 1$. (a) shows a caustic ring aligned with the galactic plane, with $a/r_0 = 15/17$ and $p/a = 1/15$. The tricusp cross section of the ring is shown in detail in (b). (c) shows a caustic ring inclined to the galactic plane. The scale shows $\log_e$[$E$/(GeV/cc)$^2$ kpc]. \label{figure6}}
\end{figure}

Consider an observer located at a distance $r_0$ from the center of a circular caustic ring with $p=q$ and $r_0 > a+p$. Fig. \ref{figure6} shows simulated skymaps of the emission measure. The caustic ring is aligned with the galactic plane in (a) and inclined to the plane in (c). The emission measure is largest when the line of sight is tangent to the ring. The tricusp shape of the cross section is clearly visible in (b). 

\subsection{Observer inside the caustic ring tube}

\begin{figure}[!h]
\begin{center}
\scalebox{0.7}{\includegraphics{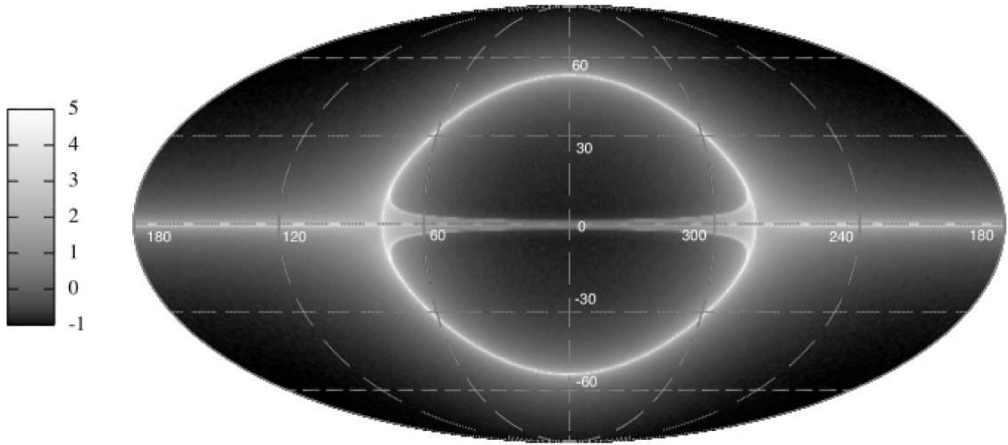}}
\end{center}
\caption{ Simulated sky map of the emission measure from an inner caustic ring, with the observer located within the tricusp tube. The ring is aligned with the galactic plane. $p/q = 1$ and $f=0.02$. The emission measure is largest when the lines of sight pass close to the cusps, resulting in the bright ring seen in the figure. The scale shows $\log_e$[E/(GeV/cc)$^2$ kpc].\label{figure7}}
\end{figure}

Let us now consider the case where the observer is located within the tricusp shaped tube of a circular caustic ring. The observer is in the plane of the ring at a distance $r_0$ from the center with $a<r_0<a+p$. We also set $p=q$. Fig. \ref{figure7} shows a simulated skymap  of the emission measure. The bright ring and the line at $b=0$ occur when the lines of sight pass close to the cusps.

We see that for the inner caustics, the emission measure can be much larger than that obtained from a smooth halo (except near the center of the halo). For a comparison of the annihilation flux with the measured gamma ray background, we refer the reader to \cite{an}.

\section{Conclusion \label{section4}}

We have discussed dark matter annihilation in outer and inner caustics and provided simulated skymaps of the emission measure. In Section \ref{section2}, we considered spherical outer caustics. When the observer is located outside the sphere, the emission measure $E$ is largest when the line of sight is tangent to the sphere, falling off rapidly to zero thereafter (Figs \ref{figure2},\ref{figure3}). When the observer is within the sphere, the emission measure only changes gradually with the angle $\psi$ (Figs. \ref{figure4},\ref{figure5}). 

In Section \ref{section3}, we considered circular, symmetric tricusp ring caustics. When the observer is outside the caustic ring, the emission measure is largest when the line of sight is tangent to the ring. Looking tangentially, one may hope to identify the tricusp cross section of the ring[Fig. \ref{figure6}(b)]. When the observer is within the tricusp, the emission measure is largest when the lines of sight pass close to the cusps, resulting in the bright ring seen in Fig. \ref{figure7}. We find that for inner caustics, $E$ can be significantly larger than that produced by a smooth halo, especially in directions away from the halo center.

\acknowledgments{This work was supported in part by the U.S. Department of Energy under contract DE-FG02-97ER41029. A.N. acknowledges financial support in part from the Deutsche Forschungsgemeinschaft(DFG) International Research Training Group GRK 881. P.S. gratefully acknowledges the hospitality of the Aspen Center for Physics while working on this project.}

\begin{appendix}

\section{ Dark matter density near an axially symmetric caustic ring. \label{app}}

Consider a single cold flow of dark matter particles falling in and out of the halo, forming a caustic. Let us consider cylindrical co-ordinates $(\rho,z)$ where $\rho = \sqrt{x^2 + y^2}$. For simplicity, we assume axial symmetry about the $z$ axis and reflection symmetry about the $z=0$ plane. Let $\rho = a$ be the caustic ring radius, defined as the point of closest approach of the particles in the $z=0$ plane (the particles with the most angular momentum are in this plane). Consider a reference sphere such that each particle in the flow passes through the sphere once. We can then assign to each particle in the flow, a two parameter label $(\alpha,\tau)$ which serves to identify the particle. We choose $\alpha = \pi/2 - \theta$ where $\theta$ is the polar angle of the particle at the time it crossed the reference sphere. $\tau = 0$ is the time when the particles just above the $z=0$ plane cross this plane. See \cite{ps_crs} for details. We may expand $\rho$ and $z$ in a Taylor series \cite{ps_crs},
\begin{eqnarray}
\rho(\alpha,\tau) - a &=& \frac{1}{2} u \left( \tau - \tau_0 \right )^2 - \frac{1}{2} s \alpha^2 \nonumber \\
z(\alpha,\tau) &=& b \, \alpha \, \tau 
\label{rhoz}
\end{eqnarray}
where $b,u,s,\tau_0$ and the ring radius $a$ are constants for a given flow. In terms of these constants, we can express the horizontal ($p$) and vertical ($q$) extents of the tricusp:
\begin{eqnarray}
p &=& \frac{1}{2} u {\tau_0}^2 \nonumber \\
q &=& \frac{\sqrt{27}}{4} \,\frac{bp}{\sqrt{us}}. 
\label{pq}
\end{eqnarray}
Let us define the dimensionless variables $X$ and $Z$
\begin{eqnarray}
X &=& \frac{\rho - a}{p} \nonumber \\
Z &=& \frac{z}{q}.
\label{XZ1}
\end{eqnarray}
Using Eq. \ref{pq} and Eq. \ref{XZ1}, we may express Eq. \ref{rhoz} in terms of $X$ and $Z$
\begin{eqnarray}
X &=& \left ( T - 1 \right )^2 - \frac{27}{64} \left( \frac{b \tau_0 \alpha}{q} \right )^2 \nonumber \\
Z &=& \frac{b \tau_0 \alpha}{q}  \, T
\label{XZ2}
\end{eqnarray}
where $T = \tau / \tau_0$. The density at points close to the caustic is given by \cite{ps_crs}
\begin{equation}
d(X,Z^2) = \frac{1}{\rho} \sum \frac{dM}{d\Omega dt} \frac{ \cos\alpha}{|D_2|} 
\label{d}
\end{equation}
where $|D_2|$ is the absolute value of the two dimensional Jacobian determinant $ \left |  \partial (x,y) / \partial (\alpha,\tau) \right | $
\begin{equation}
\left| D_2 \right | = \left | 2 b p \left[ 1 - X + 2 T^2 - 3 T \right ] \right | 
\label{D2}
\end{equation} 
and the sum is over the individual dark matter flows that exist at that point. We use the self-similar infall halo model with angular momentum \cite{ss2} to estimate the mass infall rate
\begin{equation}
\frac{dM}{d\Omega dt} = f\, v \, \frac{v^2_{rot} }{4 \pi G}.
\label{infall}
\end{equation}
$f$ is a parameter that characterizes the density of the flow , $v$ is the speed of the particles forming the caustic, and $v_{rot}$ is the rotation velocity of the halo. 

\subsection{Case 1:  $Z=0$  }

Let us first consider the simple case of points in the $Z=0$ plane. When $Z=0$, there are two flows at each point when $X > 1$, four flows at each point for $0 \le X \le 1$ and two when $X < 0$.

\subsubsection{$X > 1$}

When $X > 1$, we have two possible solutions: $T = 1 \pm \sqrt{X}$
corresponding to the two flows that exist at each point. Using 
Eq.\ref{d}, \ref{D2} and \ref{infall}, and approximating 
$\cos\alpha \approx 1$ and $a \gg p$ we find the dark matter 
density at points close to the caustic
\begin{equation}
d(X,0; X > 1) = 
\frac{f v^2_{rot}}{4 \pi G} \, \frac{v}{b} \, \frac{1}{ap} \, \frac{1}{X-1}.
\label{X1}
\end{equation}

\subsubsection{$0 \le X \le 1$}

In this range, we have four possible solutions: $T = 0,0,1 \pm \sqrt{X}$. Summing over the four solutions, we find
\begin{equation}
d(X,0; 0 \le X \le 1) = \frac{f v^2_{rot}}{4 \pi G} \, \frac{v}{b} \, \frac{1}{ap} \, \frac{1}{1-X} \left( 1 + \frac{1}{\sqrt{X}} \right ).
\label{0X1}
\end{equation}

\subsubsection{$X < 0$}

For $X < 0$, we once again have two solutions $T = 0,0$. The dark matter density is:
\begin{equation}
d(X,0; X < 0) = \frac{f v^2_{rot}}{4 \pi G} \, \frac{v}{b} \, \frac{1}{ap} \, \frac{1}{1-X}.
\label{0X}
\end{equation}

\subsection{Case 2: $Z \ne 0$}

We now consider the general case of computing the dark matter density at points outside the $Z=0$ plane. For $Z \ne 0$, neither $\alpha$ nor $T$ can be zero, so we may express $\alpha$ in terms of $T$ in Eq. \ref{XZ2}, to obtain a quartic equation for $T$:
\begin{equation}
T^4 - 2 T^3 + (1-X) T^2 - \frac{27}{64} Z^2 = 0.
\label{quartic}
\end{equation}
The dark matter density is given by
\begin{equation}
d(X,Z^2) = \frac{f v^2_{rot}}{4 \pi G}  \, \frac{v}{b} \; \sum_i \frac{1}{ 2 \left| 1 - X + 2\,T^2_i - 3 T_i  \right | }
\label{dm_density}
\end{equation}
where $T_i$ are the real roots of Eq. \ref{quartic}. The number of real roots is given by the sign of the discriminant
\begin{equation}
\beta = - \frac{27 Z^2}{4} \, \left[ \left( \frac{27 Z^2}{16} \right )^2 + \frac{27 Z^2}{16} \left ( 2 X^2 + 5 X - \frac{1}{4} \right ) + X (X - 1)^3 \right ].
\end{equation}
There are four real roots if $\beta > 0$ and two if $\beta < 0$.

To solve Eq. \ref{quartic}, let us make the substitution $T_1 = T - 1/2$, thereby eliminating the cubic term in Eq. \ref{quartic}:
\begin{eqnarray}
{T_1}^4 - \left( X + \frac{1}{2} \right ) {T_1}^2 - X T_1 - \Gamma = 0 
\label{nocubic}
\end{eqnarray}
where $\Gamma =  \frac{27}{64} Z^2 + \frac{X}{4} - \frac{1}{16}$. We may solve Eq. \ref{nocubic} by expressing it as the product of two quadratics:
\begin{equation}
\left( {T_1}^2 + C T_1 + D \right ) \left( {T_1}^2 - C T_1 - \frac{\Gamma }{D} \right ) = 0.
\end{equation}
$C^2$ solves the cubic
\begin{equation}
\left[ C^2 \right ]^3 - 2 (X + \frac{1}{2}) \left[ C^2 \right ]^2 + \left [ \left (X+\frac{1}{2} \right )^2 + 4\Gamma \right ] C^2 - X^2 = 0
\label{cubic}
\end{equation}
and
\begin{equation}
D = \frac{1}{2} \left[ C^2 - (X + \frac{1}{2}) + \frac{X}{C} \right ].
\label{D}
\end{equation}
Let us define the two variables $M$ and $N$
\begin{eqnarray}
M &=& \left( \frac{X - 1}{3} \right )^2 - \frac{9 Z^2}{16}  \nonumber \\
N &=& - \left( \frac{X-1}{3} \right )^3 - \frac{9 Z^2}{16} \left( X + \frac{1}{2} \right ).
\end{eqnarray}
$C^2$ has one real root
\begin{equation}
C^2 = \frac{2}{3} \left( X + \frac{1}{2} \right ) + \left( N + \sqrt{N^2 - M^3} \right )^{1/3} + \left( N - \sqrt{N^2 - M^3} \right )^{1/3}
\end{equation}
if $N^2 > M^3$ and three real roots 
\begin{equation}
C^2 = \frac{2}{3} \left( X + \frac{1}{2} \right ) + 2\sqrt{M} \, \cos \left( \frac{2 n \pi}{3} + \frac{1}{3} \cos^{-1} \frac{N}{M\sqrt{M}} \right )
\end{equation}
for $n = 0,1,2$ if $N^2 < M^3$. $D$ may be obtained using any of the values of $C$. The desired solution of the quartic Eq. \ref{quartic} is then
\begin{eqnarray}
T = 
  \left \{     
    \begin{array}{ll}
  \frac{1-C}{2} \pm \sqrt{ \left(\frac{C}{2}\right)^2 - D} \\
  \frac{1+C}{2} \pm \sqrt{  \left(  \frac{C}{2} \right )^2 + \frac{\Gamma}{D} }
      \end{array}.
\right.  \label{final}
\end{eqnarray}
Once the real roots $T_i$ are known, Eq. \ref{dm_density} and $v \approx b$ may be used to obtain the dark matter density.

\end{appendix}

\pagebreak

\pagebreak


\end{document}